\documentclass[10pt]{article}
\usepackage[dvips]{graphicx}  

\usepackage{color}

\begin{document}

\newcommand{\rum}{\rule{0.5pt}{0pt}}
\newcommand{\rub}{\rule{1pt}{0pt}}
\newcommand{\rim}{\rule{0.3pt}{0pt}}
\newcommand{\numtimes}{\mbox{\raisebox{1.5pt}{${\scriptscriptstyle \rum\times}$}}}
\newcommand{\numtimess}{\mbox{\raisebox{1.0pt}{${\scriptscriptstyle \rum\times}$}}}
\newcommand{\Boldsq}{\vbox{\hrule height 0.7pt
\hbox{\vrule width 0.7pt \phantom{\footnotesize T}%
\vrule width 0.7pt}\hrule height 0.7pt}}
\newcommand{\two}{$\raise.5ex\hbox{$\scriptstyle 1$}\kern-.1em/
\kern-.15em\lower.25ex\hbox{$\scriptstyle 2$}$}

\renewcommand{\refname}{References}
\renewcommand{\tablename}{\small Table}
\renewcommand{\figurename}{\small Fig.}
\renewcommand{\contentsname}{Contents}

\begin{center}
{\Large\bf  Dynamical 3-Space Predicts Hotter  Early Universe: Resolves CMB-BBN $^7$Li and $^4$He Abundance   Anomalies  \rule{0pt}{13pt}}\par

\bigskip
Reginald T. Cahill \\ 
{\small\it  School of Chemistry, Physics and Earth Sciences, Flinders University,
Adelaide 5001, Australia\rule{0pt}{15pt}}\\
\raisebox{+1pt}{\footnotesize E-mail: Reg.Cahill@flinders.edu.au}\par
\bigskip

{\small\parbox{11cm}{%

The observed abundances of   $^7Li$ and  $^4He$  are significantly inconsistent with the predictions from Big Bang Nucleosynthesis (BBN) when using the $\Lambda$CDM cosmological model together with the value for 
$\Omega_Bh^2$ $=0.0224\pm0.0009$ from WMAP CMB fluctuations,  with the value from BBN required to fit observed abundances  being $0.009<\Omega_Bh^2<0.013$.  The dynamical 3-space theory is shown to predict a 20\% hotter universe in the radiation-dominated epoch, which then results in a remarkable parameter-free agreement between the BBN and the WMAP value for $\Omega_Bh^2$. The dynamical 3-space also gives a parameter-free fit to the supernova redshift data, and predicts that the flawed $\Lambda$CDM  model would require $\Omega_\Lambda=0.73$ and $\Omega_M=0.27$ to fit the 3-space dynamics Hubble expansion, and independently of the supernova data. These results amount to the discovery of new physics for the early universe that is matched by numerous other successful observational and experimental tests.
 
 \rule[0pt]{0pt}{0pt}}}\medskip
\end{center}

\setcounter{section}{0}
\setcounter{equation}{0}
\setcounter{figure}{0}
\setcounter{table}{0}

\markboth{R.T.  Cahill.  Dynamical 3-Space Predicts Hotter Early Universe: Resolves CMB-BBN $^7$Li and $^4$He  Abundance   Anomalies  }{\thepage}
\markright{R.T.  Cahill. Dynamical 3-Space Predicts Hotter early Universe:  Resolves CMB-BBN$^7$Li and $^4$He Abundance   Anomalies }

\tableofcontents

\section{Introduction}

Astrophysical observed abundances of  $^7Li$  and $^4He$ are significantly inconsistent with the predictions from Big Bang Nucleosynthesis (BBN) when using the $\Lambda$CDM cosmological model, with the value for\footnote{$H_0=100 h$ km/s/Mpc defines $h$.  $\Omega_B$ is baryon density relative to critical density $\rho_c$.}  
$\Omega_Bh^2$ $=0.0224\pm0.0009$ from WMAP CMB fluctuations being considerably different from the value from BBN required to fit observed abundances  $0.009<\Omega_Bh^2<0.013$  Coc {\it et al.} \cite{Coc1}. 

The most significant long-standing discrepancy is that of $^7$Li because the pre-Galactic lithium abundance   inferred from observations of metal-poor (Population II) stars is at least 2-3 times  smaller than predicted by BBN--$\Lambda$CDM. The $^7$Li problem has been most difficult to understand as its primordial abundance  should be the most reliable, because of the higher observational statistics and an easier extrapolation to primordial values. Various possible resolutions were discussed in \cite{Cybert}, with the conclusion that the lithium problem most likely points to new physics.

It is shown herein that the new physics of a dynamical 3-space \cite{Review, Book, Paradigm} results in a 20\% hotter universe during the radiation dominated epoch, and in a parameter-free analysis the BBN abundances are brought into close agreement with the WMAP value for  the baryonic density $\Omega_Bh^2$ $=0.0224\pm0.0009$. The dynamical 3-space also gives a parameter free account of the  supernova redshift data, and  fitting the $\Lambda$CDM to the dynamical 3-space model requires $\Omega_\Lambda=0.73$ and $\Omega_m=0.27$, independently of the supernova data.  There are numerous other experimental and observational confirmations of the new physics \cite{Book,Review}, including a recent analysis of the NASA/JPL spacecraft earth-flyby Doppler-shift  anomalies  \cite{And2008, CahillNASA}. The conclusion is that the $\Lambda$CDM is flawed, with precision data from the supernova redshifts \cite{PS,S1,S2},  and WMAP CMB fluctuations \cite{WMAP} in conjunction with BBN computations finally ruling out this model.  As briefly noted below that $\Lambda$CDM is essentially Newtonian gravity, and various data have indicated the failure of Newtonian gravity.   

\section{Dynamical 3-Space}

Newton's inverse square law of gravity \cite{Newton} has the differential form
\begin{equation}
\nabla.{\bf g}=-4\pi G\rho,  \mbox{\ \ \  } \nabla \times {\bf g} = {\bf 0}, 
\label{eqn:Newton}\end{equation}
for the matter acceleration field ${\bf g}({\bf r},t)$.
Application of this to spiral galaxies and the expanding universe has lead to many problems, including, in part, the need to invent dark energy and dark matter.  However (\ref{eqn:Newton}) has a unique generalisation that resolves these problems.   In terms of a velocity field ${\bf v}({\bf r},t)$ (\ref{eqn:Newton}) has an equivalent form \cite{Book,Review}
\begin{equation}
\nabla.\left(\frac{\partial {\bf v} }{\partial t}+ ({\bf v}.{\bf \nabla}){\bf v}\right)
=-4\pi G\rho,  \mbox{\ \ \  } \nabla \times {\bf v} = {\bf 0}, 
\label{eqn:Newtonv} \end{equation}
where now
\begin{equation}
{\bf g}=\frac{\partial{\bf v}}{\partial t}+({\bf v}.\nabla){\bf v},
\label{eqn:acceln}\end{equation}
is the Euler acceleration of the substratum that has velocity  ${\bf v}({\bf r},t)$.
Because of the covariance of $v$ under a change of the spatial coordinates only relative internal velocities have an ontological existence - the coordinates ${\bf r}$  then merely define a mathematical embedding space.
In the form  (\ref{eqn:Newtonv}) Newton's law permits a unique generalisation by adding a term of the same order but which can preserve the inverse square law outside of spherical masses,
\begin{eqnarray}
\nabla.\left(\frac{\partial {\bf v} }{\partial t}+ ({\bf v}.{\bf \nabla}){\bf v}\right)+
\frac{\alpha}{8}\left((tr D)^2 -tr(D^2)\right)=-4\pi G\rho,  \nonumber 
\label{eqn:vflow}\end{eqnarray}
\vspace{-6mm}
\begin{eqnarray}
 \nabla\times {\bf v}={\bf 0},  \mbox{\  \  \   }
 D_{ij}=\frac{1}{2}\left(\frac{\partial v_i}{\partial x_j}+
\frac{\partial v_j}{\partial x_i}\right).
\label{eqn:3spacedynamics}\end{eqnarray} 
Eqn  (\ref{eqn:vflow}) has two fundamental constants: $G$ and $\alpha$.  Experimental bore-hole $g$ anomaly data reveals that $\alpha$ is the fine structure constant $\approx 1/137$ to within experimental errors \cite{Review,Book}.  Eqn  (\ref{eqn:vflow}) has a rich variety of  solutions: (i) black holes with a non-inverse square law acceleration field that explains the supermassive black hole  mass spectrum and the flat rotation curves of spiral galaxies without the need for dark matter - these black holes may be primordial as well as induced, (ii) the bore-hole g-anomaly, (iii) gravitational light  bending,  (iv) a  parameter free fit to the supernova data \cite{Paradigm} without the need for dark energy or dark matter, and other effects. As well the 3-space field ${\bf v}({\bf r},t)$ has been directly detected in numerous laboratory experiments, and now in Doppler shift data from spacecraft earth-flybys \cite{CahillNASA}.

Eqn  (\ref{eqn:vflow})  gives a different account of the  Hubble expansion of the universe, and here we outline a new account of the thermal history of the universe.   The results are very different from the predictions of the Friedmann equation - the standard equation of cosmology since its inception (FRW-GR). In the Friedmann equations the expansion of the universe is determined solely by the presence of matter or energy, as would be expected since it derives from (\ref{eqn:Newton}), and it then requires, at the present epoch, some  $73\%$ dark energy, $23\%$ dark matter  and $4\%$ baryonic matter.  Eqn  (\ref{eqn:vflow}), in contrast,  requires only the normal matter - this is because  (\ref{eqn:vflow})  has an expanding 3-space solution even in the absence of matter/energy. Fitting the Friedmann Hubble function $H(z)$ to the Hubble function from (\ref{eqn:vflow}), using the usual distance-redshift modulus as a measure,  indeed permits these dark energy and dark matter quantities to be simply predicted, independently of the observed supernova data, for these are the values that best-fit the $\Lambda$CDM to  the observed uniformly expanding 3-space Hubble solution.

\begin{figure}
\hspace{3mm}\includegraphics[scale=0.5]{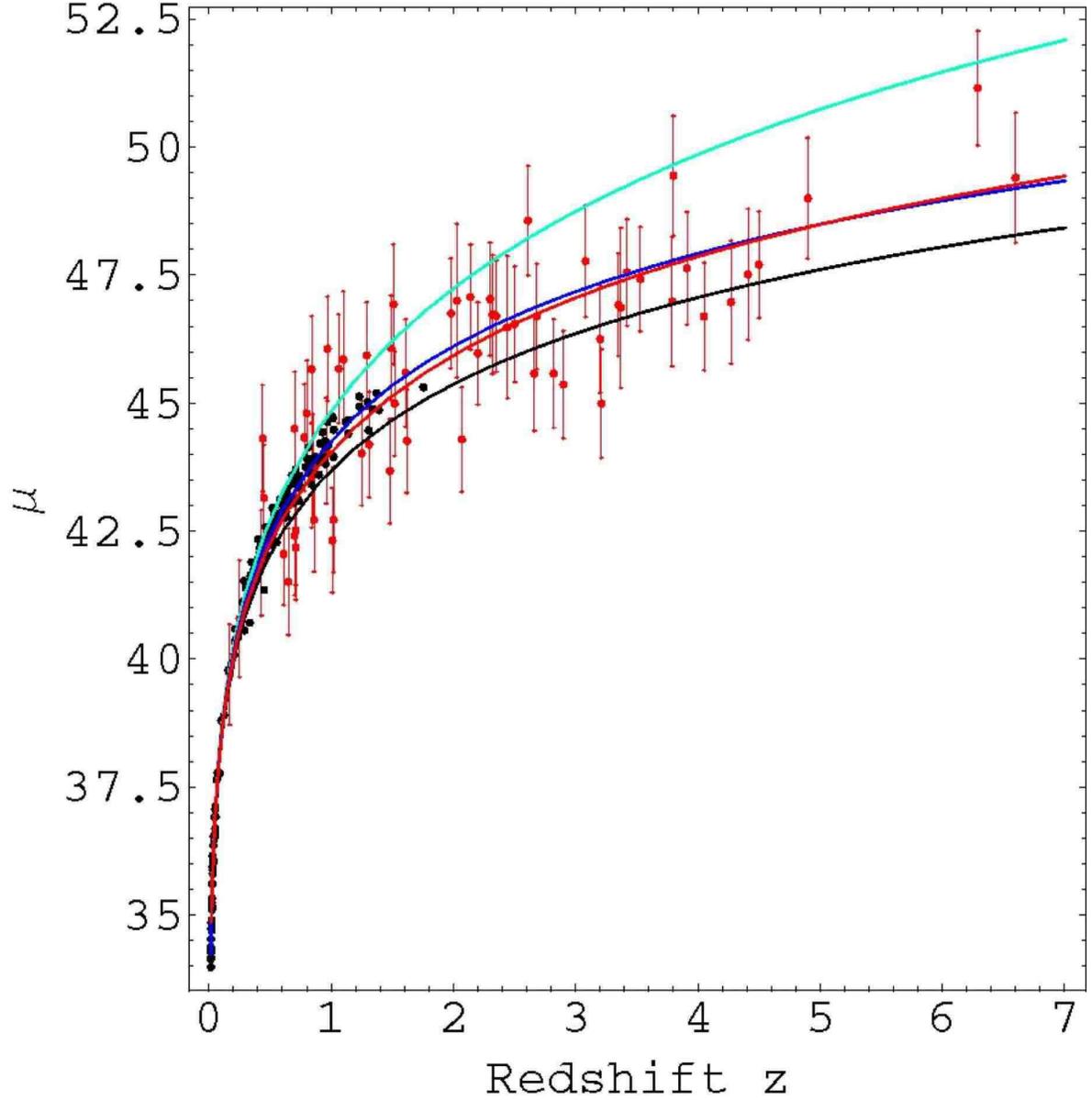}
\vspace{-4mm}\caption{\small{ Hubble diagram showing the supernovae data  using several data sets, and the Gamma-Ray-Bursts data (with error bars).  Upper curve (green)  is $\Lambda$CDM `dark energy' only $\Omega_\Lambda=1$, lower curve  (black) is $\Lambda$CDM matter only $\Omega_M=1$. Two middle curves show best-fit of $\Lambda$CDM `dark energy'-`dark-matter' (blue) and dynamical 3-space prediction (red), and are essentially indistinguishable.   We see that the best-fit  $\Lambda$CDM `dark energy'-`dark-matter' curve essentially converges on the uniformly-expanding parameter-free dynamical 3-space prediction.  The supernova data  shows that the universe is undergoing a uniform expansion, although not reported as such in \cite{S1,S2,PS}}, wherein a fit to the FRW-GR  expansion was forced, requiring `dark energy', `dark  matter' and a future `exponentially accelerating expansion'.
\label{fig:SN1}}\end{figure}

\section{Expanding Universe from Dynamical 3-Space}

Let us now explore the expanding  3-space  from (\ref {eqn:vflow}).  Critically, and unlike the FLRW-GR model, the 3-space expands even when the energy density is zero.
Suppose that  we have a radially symmetric effective density $\rho(r,t)$, modelling normal matter and EM radiation,  and that we look for a radially symmetric time-dependent flow ${\bf v}({\bf r},t)$$ =v(r,t)\hat{\bf r}$ from (\ref{eqn:vflow}).  Then $v(r,t)$ satisfies the equation,  with $v^\prime=\displaystyle{\frac{\partial v(r,t)}{\partial r}}$,
 \begin{eqnarray}
\lefteqn{\frac{\partial}{\partial t}\left( \displaystyle{\frac{2v}{r}}+v^\prime\right)+ vv^{\prime\prime}+2\frac{vv^{\prime}}{r}+ (v^\prime)^2 } \nonumber  \\
& & + \frac{\alpha}{4}\left(\frac{v^2}{r^2} +\frac{2vv^\prime}{r}\right)=- 4\pi G \rho(r,t). 
 \label{eqn:radialflow}\end{eqnarray}
 Consider first the zero energy case $\rho=0$. Then we have a Hubble  solution $v(r,t)=H(t)r$, a centreless flow, determined by
\begin{equation}{\dot H}+\left(1+\frac{\alpha}{4}\right)H^2=0,
\end{equation}
with ${\dot H}=\displaystyle{\frac{dH}{dt}}$.  We also introduce in the usual manner the scale factor $a(t)$ according to $H(t)=\displaystyle{\frac{\dot{a}}{a}}$. We then obtain
the solution
\begin{equation}
H(t)=\frac{1}{(1+\frac{\alpha}{4})t}=H_0\frac{t_0}{t}; \mbox{\ \  }  a(t)=a_0\left(\frac{t}{t_0} \right)^{4/(4+\alpha)}
\label{eqn:spacexp}\end{equation}
where $H_0=H(t_0)$ and $a_0=a(t_0)=1$, with $t_0$ the present age of the universe.  Note that we obtain an expanding 3-space even where the energy density is zero - this is in sharp contrast to the FLRW-GR model for the expanding universe, as shown below.
The solution (\ref{eqn:spacexp}) is unique - it has one free parameter - which is essentially the age of the universe $t_0=t_H=1/H_0$, and clearly this cannot be predicted by physics, as it is a purely contingent effect - the age of the universe when it is observed by us.  Below we include the small effect of ordinary matter and EM radiation.

We can write the  Hubble function $H(t)$ in terms of $a(t)$ via the inverse function $t(a)$, i.e. $H(t(a))$ and finally as $H(z)$, where the redshift observed now, relative to the wavelengths at time $t$, is  $z=a_0/a-1$. Then we obtain
\begin{equation}
H(z)={H_0}(1+z)^{1+\alpha/4}
\label{eqn:H2a}\end{equation}
To test this expansion we need to predict the relationship between the cosmological observables, namely the apparent photon energy-flux magnitudes and redshifts. This  involves taking account of the reduction in photon count caused by the expanding 3-space, as well as the accompanying reduction in photon energy.  The result is that the dimensionless `energy-flux'  luminosity effective distance is then given by
 \begin{equation}
d_L(z)=(1+z)\int_0^z \frac{H_0 dz^\prime}{H(z^\prime)}
\label{eqn:H1a}\end{equation}
 and the  distance modulus is  defined as usual by
\begin{equation}
\mu(z)=5\log_{10}(d_L(z))+m.
\label{eqn:H1b}\end{equation}
Because all the selected supernova have the same absolute magnitude, $m$ is a constant whose value is determined by fitting the low $z$ data. 

Using the  Hubble expansion (\ref{eqn:H2a}) in (\ref{eqn:H1a}) and (\ref{eqn:H1b}) we obtain the middle curve (red) in Fig.\ref{fig:SN1}, yielding an excellent agreement with the supernovae and GRB data. Note that because $\alpha/4$ is so small it actually has negligible effect on these plots.  But that is only the case for the homogeneous expansion - the $\alpha$ dynamics can result in large effects such as black holes and large spiral galaxy rotation effects when the 3-space is inhomogeneous, and particularly precocious galaxy formation.  Hence the dynamical 3-space gives an immediate account of the universe expansion data, and does not require the introduction of  a cosmological constant or `dark energy' nor `dark matter'.

 \begin{figure}
\vspace{-12mm}
\hspace{25mm}\includegraphics[scale=1.3]{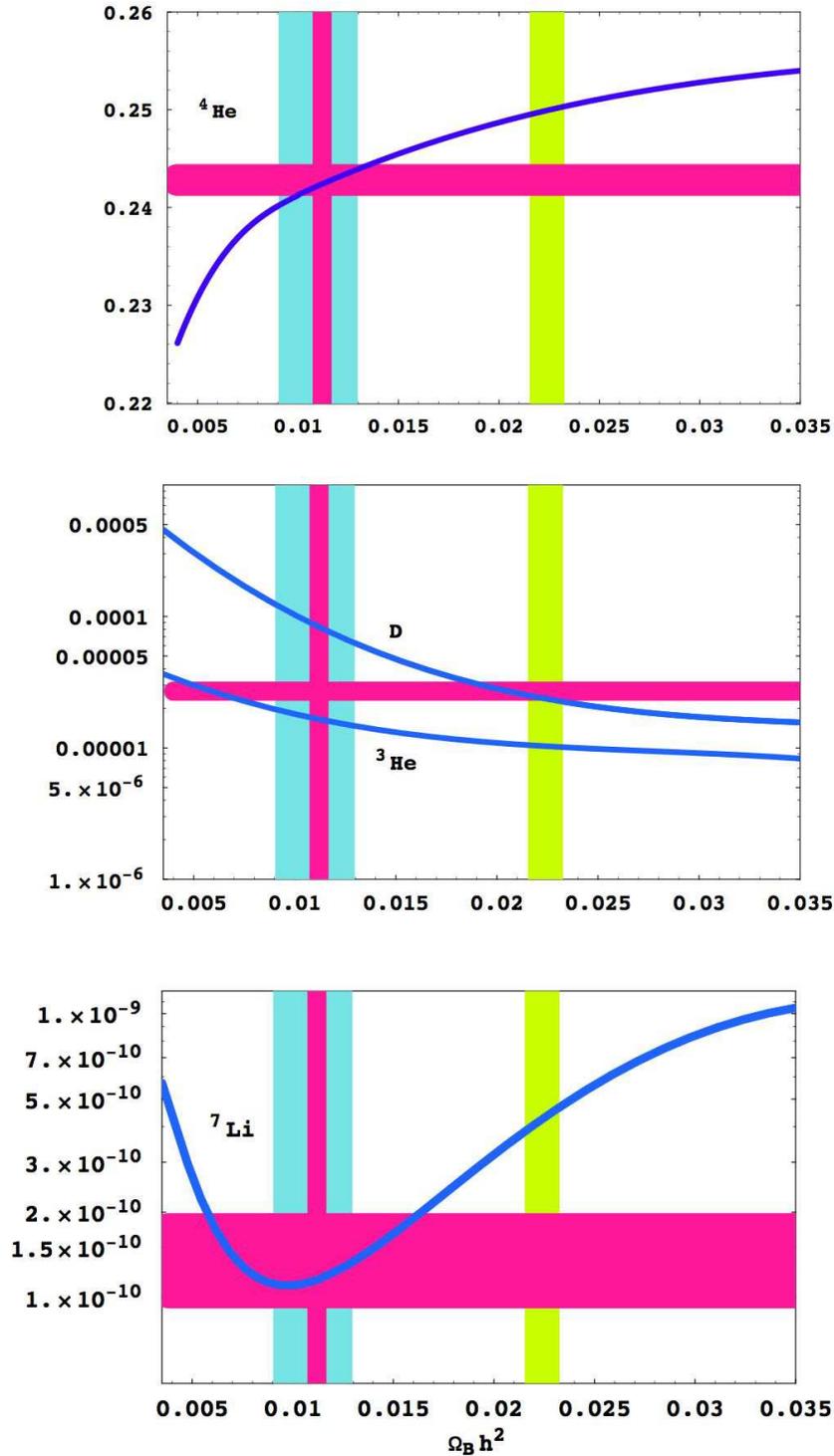}
\vspace{-5mm}\caption{\small{Shows the Big Bang nucleosynthesis (BBN) number abundances for:  the$^4$He mass fraction (top), D and $^3$He (middle) and $^7$Li (bottom)  relative to hydrogen vs $\Omega_Bh^2$, as blue curves, from Coc {\it et al.}\cite{ Coc1}.
Horizontal (red) bar-graphs show astrophysical abundance observations.  The vertical (yellow) bar-graphs show the values $\Omega_Bh^2$ $=0.0224\pm0.0009$ from WMAP CMB fluctuations, while the (blue) bar-graph $0.009<\Omega_Bh^2<0.013$ shows the best-fit at 68\% CL from the BBN for the observed abundances \cite{Coc1}.  We see that the WMAP data is in significant disagreement with the BBN results for $\Omega_Bh^2$, giving, in particular, the $^7$Li abundance anomaly within the $\Lambda$CDM model.  The dynamical 3-space model has a different and hotter thermal history in the radiation dominated epoch, and the corresponding BBN predictions are  easily obtained by a  re-scaling of the WMAP value $\Omega_Bh^2$ to $\Omega_Bh^2/2$. The  resultant 
$\Omega_Bh^2$  $=0.0112\pm 0.0005$ values are shown by the vertical (red) bar-graphs that center on the BBN $0.009<\Omega_Bh^2<0.013$  range, and which is now in remarkable agreement with BBN computations. So while the BBN - WMAP inconsistency indicates a failure of the Friedmann FRW-GR Big Bang model, it is another success for the new physics entailed in the dynamical 3-space model.  Plots adapted from \cite{Coc1}.}
\label{fig:BBN}}\end{figure}

\section{Expanding Universe - Matter and Radiation Only}
When the energy density is not zero we need to take account of the dependence of $\rho(r,t)$ on the scale factor of the universe. In the usual manner we thus write
\begin{equation}
\rho(r,t)=\frac{\rho_{m}}{a(t)^3}+\frac{\rho_{r}}{a(t)^4},
\label{eqn:evolve}\end{equation}
for ordinary matter and EM radiation. Then (\ref{eqn:radialflow}) becomes for $a(t)$
\begin{eqnarray}
\frac{\ddot  a}{a}+\frac{\alpha}{4}\frac{{\dot a}^2}{a^2}
=-\frac{4\pi G}{3}\left(\frac{\rho_{m}}{a^3}+\frac{\rho_{r}}{a^4} \right),
\label{eqn:Reqn}\end{eqnarray}
giving
\begin{equation}
{\dot a}^2=\frac{8\pi G}{3}\left(\frac{\rho_{m}}{a}+\frac{\rho_{r}}{2a^2}\right)-\frac{\alpha}{2}\int\frac{{\dot a}^2}{a}da+f,
\label{eqn:R2}\end{equation}
where $f$ is the integration constant.
 In terms of ${\dot a}^2$ this has the solution
\begin{equation}
{\dot a}^2\!=\!\frac{8\pi G}{3}\left(\!\frac{\rho_{m}}{(1-\frac{\alpha}{2})a}\!+\!\frac{\rho_{r}}{(1-\frac{\alpha}{4})2a^2}+b a^{-\alpha/2}\!\right),
\label{eqn:R3}\end{equation}
which is easily checked by substitution into (\ref{eqn:R2}), and where $b$ is the  integration constant.  We have written an overall factor of $8\pi G/3$ even though b, in principle, is independent of $G$. This gives $b$ convenient units of matter density, but which does not correspond to any actual energy.  From now on we shall put $\alpha=0$. Finally we obtain from  (\ref{eqn:R3})
\begin{equation}
t(a)=\int^a_0\frac{da}{\sqrt{\displaystyle{\frac{8\pi G}{3}}\left(\displaystyle{\frac{\rho_{m}}{a}+\frac{\rho_{r}}{2a^2}}+b \right)}}.
\label{eqn:R4}\end{equation} 
 When $\rho_m=\rho_r=0$, (\ref{eqn:R4})
reproduces the expansion in (\ref{eqn:spacexp}), and so the density terms in (\ref{eqn:R3}) give the modifications to  the dominant purely-spatial expansion, which we have noted above  already gives an excellent account of the red-shift data. Having $b\neq 0$ simply asserts that the 3-space can expand even when the energy density is zero - an effect missing from FLRW-GR cosmology.
From (\ref{eqn:R3})  we  obtain\footnote{From now-on an `overline' is used to denote the 3-space   values. Note that $\overline{H}_0 \equiv H_0$ - the current observable value.}
\begin{equation}
\overline{H}(z)^2={H_0}^2(\overline{\Omega}_m(1+z)^3+\frac{\overline{\Omega}_r(1+z)^4}{2} +\overline{\Omega}_s(1+z)^{2}),
\label{eqn:H2}\end{equation}
\begin{equation}
\overline{\Omega}_m \equiv \rho_m/\rho_c,  \mbox{\ \ \  }\overline{\Omega}_r \equiv  \rho_r/\rho_c, \mbox{\ \ \  }\overline{\Omega}_s \equiv  b/\rho_c,
\end{equation}
\begin{equation}
\overline{\Omega}_m+\frac{\overline{\Omega}_r}{2}+\overline{\Omega}_s=1.
\end{equation}
\begin{equation}
H_0=\left(\frac{8\pi G}{3}(\rho_m+\frac{\rho_r}{2}+b)\right)^{1/2}\equiv\left(\frac{8\pi G}{3}\rho_c\right)^{1/2}
\label{eqn:Hubbleconstant}\end{equation}
which defines the usual critical energy density $\rho_c$, but which here is merely a form for $H_0$ - it has no interpretation as an actual energy density, unlike in FRW-GR. 
Note the factor of 2 for $\overline{\Omega}_r$, which is a key effect in this paper, and is not in FRW-GR. In the dynamical 3-space model these $\overline{\Omega}$'s do not correspond  to the composition of the universe, rather to the relative dynamical effects  of the matter and radiation on the intrinsic 3-space expansion dynamics.
 $H_0= 73 $ km/s/Mpc with $\overline{\Omega}_m\approx \Omega_B =0.04$ and 
$\overline{\Omega}_s=0.96$ gives an age for the universe of $t_0=12.6$Gyrs, while (\ref{eqn:Friedmannage}) with  $\Omega_M=0.27$ and $\Omega_\Lambda=0.73$ gives $t_0=13.3$Gyrs.
$\overline{\Omega_r}=\Omega_r=8.24\times10^{-5}$.

\section{Friedmann-GR Standard $\Lambda$CDM Cosmology Model}
We now discuss  the strange feature of the  standard model dynamics which requires a non-zero energy density for the universe to expand.   The well known Friedmann equation is
\begin{eqnarray}
\left(\frac{\dot  a}{a}\right)^2=\frac{4\pi G}{3}\left(\frac{\rho_{M}}{a^3}+\frac{\rho_{r}}{a^4}+\Lambda \right),
\label{eqn:Friedmann}\end{eqnarray}
where now $\rho_M=\rho_m+\rho_{DM}$ is the energy  composition of the universe, and includes ordinary matter and dark matter, and $\Lambda$ is the cosmological constant or dark energy, expressed in mass density units. The differences between (\ref{eqn:R2}) and (\ref{eqn:Friedmann}) need to be noted:  apart from the $\alpha$ term (\ref{eqn:Friedmann}) has no integration constant which corresponds to a purely spatial expansion, and in compensation requires the {\it ad hoc}  dark matter and dark energy terms, whose best-fit values are easily predicted; see below. 
It is worth noting how (\ref{eqn:Friedmann}) arises from Newtonian gravity.  For radially expanding homogeneous matter (\ref{eqn:Newton}) gives for the total energy $E$ of a test mass (a galaxy) of mass $m$
\begin{equation}
\frac{1}{2}mv^2-\frac{G m M(r)}{r}=E,
\label{eqn:NG}\end{equation}
where $M(r)$ is the time-independent amount of matter  within a sphere of radius $r$. With $E=0$  and $M(r)=\frac{4}{3}\pi r^3 \rho(t)$ and $\rho(t) \sim 1/r(t)^3$ (\ref{eqn:NG}) has the Hubble form  $v=H(t) r$. In terms of $a(t)$ this gives (\ref{eqn:Friedmann}) after an {\it ad hoc} and invalid inclusion of the radiation and dark energy terms, as for these terms $M(r)$ is not independent of time, as assumed above.  These terms are usually included on the basis of the  Weyl expression for the stress-energy tensor within GR.
Eqn.(\ref{eqn:Friedmann}) leads to the analogue of (\ref{eqn:R4}),
\begin{equation}
t(a)=\int^a_0\frac{da}{\sqrt{\displaystyle{\frac{8\pi G}{3}}\left(\displaystyle{\frac{\rho_{M}}{a}+\frac{\rho_{r}}{a^2}+\Lambda a^2}\right)}},
\label{eqn:Friedmannage}\end{equation} 
\begin{equation}
H(z)^2=H_0^2(\Omega_M(1+z)^3+\Omega_r(1+z)^4 +\Omega_\Lambda(1+z)^2),
\label{eqn:FH2}\end{equation}
\begin{equation}
\Omega_M \equiv \rho_M/\rho_c,\mbox{\ \ \  } \Omega_r  \equiv \rho_r/\rho_c, \mbox{\ \ \  }\Omega_\Lambda \equiv \Lambda/\rho_c,
\label{FriedmannOmega1} \end{equation}
\begin{equation}
\Omega_M+\Omega_r+\Omega_\Lambda=1.
\label{FriedmannOmega2} \end{equation}
\begin{equation}
H_0=\left(\frac{8\pi G}{3}(\rho_M+\rho_r+\Lambda)\right)^{1/2}\equiv\left(\frac{8\pi G}{3}\rho_c\right)^{1/2}.
\label{eqn:FHubbleconstant}\end{equation}
This has the same value of $\rho_c$ as in (\ref{eqn:Hubbleconstant}), but now interpreted as an actual energy density.  
Note that $\overline{\Omega}_r=\Omega_r$, but that  $\overline{\Omega}_m \neq\Omega_M$, as $\Omega_M$ includes the spurious `dark matter'. 

\section{Predicting the $\Lambda$CDM  Parameters  $\Omega_\Lambda$ and $\Omega_{DM}$}

The `dark energy' and `dark matter' arise in the FLRW-GR cosmology because in that model space cannot expand unless there is an energy density present in the space, if that space is flat and the energy density is pressure-less. Then essentially fitting the Friedmann  model  $\mu(z)$ to the dynamical 3-space cosmology $\mu(z)$ we obtain 
 $\Omega_\Lambda=0.73$, and so $\Omega_M=1-\Omega_\Lambda=0.27$. These values arise from a best fit for $z\in\{0,14\}$ \cite{Paradigm}. The actual values for $\Omega_\Lambda$ depend on the red-shift range used, as the Hubble functions for the FLRW-GR and dynamical 3-space have different functional dependence on $z$. These values are of course independent of the actual observed redshift data.  Essentially the current standard model of cosmology $\Lambda$CDM  is excluded from modelling a uniformly expanding dynamical 3-space, but by choice of the parameter $\Omega_\Lambda$ the $\Lambda$CDM Hubble function $H(z)$ can be made to best-fit the data. However $H(z)$ has the wrong functional form; when applied to the future expansion of the universe the Friedmann dynamics  produces a spurious  exponentially expanding universe.

 \section{Dynamical 3-Space and Hotter Early Universe}
 
 The 3-space dynamics and the $\Lambda$CDM dynamics give different accounts of the expansion of the universe and in particular of the thermal history during the radiation dominated epoch.
$\Lambda$CDM gives in that epoch, from the Friedmann equation (\ref{eqn:Friedmann}), $a(t)=\sqrt{2H_0t\sqrt{\Omega_r}}$, while
(\ref{eqn:R4}) gives $a(t)=$ $ \sqrt{2H_0t\sqrt{\Omega_r/2}}$. Because the CMB is thermal radiation its temperature varies as $T(t)=(2.725\pm0.001)/a(t)$  $^\circ$K, and so the 3-space dynamics predicts an early  thermal history that is 20\% hotter.  This means that a re-analysis of the BBN is required.  However this is easily achieved by a scaling analysis.  Essentially we can do this by effectively using $H_0/\sqrt{2}$ in place of $H_0$ in the radiation-dominated epoch, as this takes account of the $\Omega_r/2$ effect. In terms of  $\Omega_Bh^2$, which determines the BBN, this amounts to the re-scaling  $\Omega_Bh^2 \rightarrow \Omega_Bh^2/2$. This immediately brings the WMAP  $\Omega_Bh^2$ $=0.0224\pm0.0009$ down to, effectively, $\Omega_Bh^2$ $=0.0112\pm0.0005$, and into excellent agreement with the BBN value  $0.009<\Omega_Bh^2<0.013$, as shown in Fig.\ref{fig:BBN}, and discussed in detail in the figure caption.

\section{Conclusions}

It has been shown that the significant inconsistency between observed abundances of   $^7Li$ and  $^4He$  with the predictions from Big Bang Nucleosynthesis (BBN) when using the $\Lambda$CDM cosmological model together with the value for 
$\Omega_Bh^2$ $=0.0224$ $\pm0.0009$ from WMAP CMB fluctuations,  with the value from BBN required to fit observed abundances  being $0.009<\Omega_Bh^2<0.013$, are resolved with remarkable precision  by using the dynamical 3-space theory. This theory is shown to predict a 20\% hotter universe in the radiation-dominated epoch, which then results in a remarkable agreement between the BBN and the WMAP value for $\Omega_Bh^2$. The dynamical 3-space also gives a parameter-free fit to the supernova redshift data, and predicts that the flawed $\Lambda$CDM  model would require $\Omega_\Lambda=0.73$ and $\Omega_M=0.27$ to fit the 3-space dynamics Hubble expansion, and independently of the supernova data. These results amount to the discovery of new physics for the early universe.  This new physics has also explained (i) the bore-hole $g$ anomaly, (ii) black-hole mass spectrum, (iii) flat rotation curves in spiral galaxies, (iv) enhanced light bending by galaxies, (v)  anomalies in laboratory measurements of $G$,  (vi) light speed anisotropy experiments including the explanation of the Doppler shift anomalies in spacecraft earth-flybys, and (vii) the detection of so-called gravitational waves.   As well because (\ref{eqn:3spacedynamics}) is non-local it can overcome the horizon problem. The new physics unifies cosmology with laboratory based phenomena, indicating a new era of precision studies of the cosmos.

\end{document}